\documentclass[a4paper,11pt]{article}

\usepackage[utf8]{inputenc}
\usepackage{geometry}
\usepackage{amsmath, amssymb, amsfonts}
\usepackage{graphicx}
\usepackage{booktabs}
\usepackage{hyperref}
\usepackage{authblk}
\usepackage{setspace}

\usepackage{float}      
\usepackage[section]{placeins} 

\usepackage[superscript]{cite} 
\usepackage[small,compact]{titlesec}
\usepackage[font=small,labelfont=bf,skip=2pt]{caption}

\hypersetup{
    colorlinks=true,
    linkcolor=blue,
    filecolor=magenta,      
    urlcolor=cyan,
    citecolor=black,
}

\title{\textbf{Investigation of Toroidal Rotation Effects on Spherical Torus Equilibria using the Fast Spectral Solver VEQ-R}}

\author[1,2]{Xingyu Li}
\author[2]{Hua-sheng Xie\textsuperscript{\textdagger}}
\author[1]{Lai Wei}
\author[1]{Zheng-Xiong Wang\textsuperscript{\textdagger}}

\affil[1]{\small Key Laboratory of Materials Modification by Laser, Ion, and Electron Beams of the Ministry of Education, School of Physics, Dalian University of Technology, Dalian 116024, China}
\affil[2]{\small ENN Science and Technology Development Co., Ltd., Langfang 065001, China}

\date{}

\begin{document}

\maketitle

{
  \renewcommand{\thefootnote}{\textdagger}
  \footnotetext{Corresponding authors. E-mail: huashengxie@gmail.com (H. Xie), zxwang@dlut.edu.cn (Z.-X. Wang).}
}

\begin{abstract}
Standard reduced models often fail to adequately describe the complex geometric response of tokamak plasmas to strong toroidal rotation. In this work, we present VEQ-R, a computationally efficient spectral solver designed to calculate fixed-boundary equilibria with arbitrary toroidal flow. In contrast to computationally intensive grid-based codes, our model employs a 12-parameter shifted Chebyshev spectral expansion to explicitly resolve radial variations in high-order shaping profiles—such as dynamic elongation and triangularity. This capability allows the solver to accurately capture differential flux surface distortions (non-rigid effects) even in challenging sonic regimes ($M \sim 1.0$). By synergizing this compact variational formulation with a novel ``Matrix-Kernel'' acceleration technique, we transform the problem into pre-computed algebraic matrix operations. This approach achieves convergence in approximately 5 ms, maintaining exceptional geometric fidelity compared to high-resolution benchmarks while balancing speed and accuracy. Our analysis reveals that rotation-induced flux compression leads to a monotonic decrease in the core safety factor $q_0$, pushing it dangerously close to unity—a structural deformation mechanism effectively captured by this approximate yet robust solver.
\end{abstract}

\section{Introduction}

High-performance tokamak scenarios, such as the H-mode, are rarely static; they are characterized by strong toroidal rotation, often driven by neutral beam injection (NBI)\cite{Rice2016}. In spherical tokamaks or next-step devices, core Mach numbers can easily exceed 0.5. While these shear flows are beneficial for suppressing resistive wall modes (RWMs)\cite{Bondeson1994}, neoclassical tearing modes\cite{Hender2004}, and turbulence\cite{Burrell1997}, the associated centrifugal forces fundamentally alter the equilibrium. They push plasma pressure toward the low-field side (LFS), inducing significant magnetic axis shifts and distorting the flux surface geometry. Accurately modeling these effects is not just an academic exercise—it is essential for predicting stability boundaries and designing safe operational limits.

The numerical challenge here is substantial. The equilibrium state is governed by the Generalized Grad-Shafranov (GGS) equation\cite{Maschke1980,Hameiri1983}. High-fidelity solvers generally fall into two categories: standard fixed-grid codes like EFIT\cite{Lao1985} which rely on Picard iterations, and inverse flux coordinate solvers such as the recent work by Feng et al.\cite{Feng2024}. More recently, Chen et al.\cite{Chen2022} have extended the analysis to free-boundary equilibria, investigating how toroidal rotation significantly impacts the separatrix topology and plasma shape. While these methods provide exquisite detail, they typically require seconds or even minutes to converge due to complex grid generation and iterative schemes. This latency makes them impractical for millisecond-scale applications like real-time Plasma Control Systems (PCS) , massive integrated transport simulations, or system code level parameter scans.\cite{Chen2026}.

To bridge the gap between fidelity and speed, variational moment methods were developed to reduce the complexity of the problem. This approach was pioneered by Lao et al.\cite{Lao1981}, who reduced the partial differential equation (PDE) to a set of coupled ordinary differential equations (ODEs). To further improve computational efficiency, Haney et al.\cite{Haney1989,Haney1995} and Varadarajan et al.\cite{Varadarajan1991} formalized a parametric approach, reducing the problem to a system of algebraic equations suitable for rapid equilibrium calculation. However, these early algebraic models were typically limited by their low-order geometric approximations. They often retained only the lowest-order harmonics and assumed ``rigid'' profiles, which fail to capture the complex, non-rigid geometric distortions excited by strong centrifugal forces in sonic regimes.

Following the algebraic framework but seeking higher fidelity, we propose a fast spectral solver designed for real-time applications. Informed by recent theoretical studies by Xie and Li\cite{Xie2026} on the optimal parameterization efficiency, we extend the spectral moment framework to explicitly include toroidal flow effects. We designate this extended fixed-boundary solver as VEQ-R (where ``R'' denotes Rotation). Unlike strictly high-fidelity inverse coordinate solvers that pursue numerical exactness at the cost of computation time, VEQ-R is designed as an approximate solver that prioritizes computational speed and ease of use. By introducing a 12-degree-of-freedom spectral expansion and a novel ``Matrix-Kernel'' technique, it reduces the non-linear problem to highly optimized algebraic matrix operations.

Distinct from previous low-order attempts, our approach is defined by three synergistic innovations. First, we move beyond basic displacement and elongation, explicitly incorporating high-order triangularity terms to capture the subtle geometric response to centrifugal forces. Second, we replace the runtime numerical integration of the variational functional with pre-calculated projection matrices. This ``Matrix-Kernel'' scheme transforms the non-linear problem into highly optimized algebraic matrix operations. Consequently, implemented in a vectorized MATLAB environment, the solver converges in roughly 5 ms on a single core—offering a speedup factor of nearly 1000 compared to high-resolution finite difference codes. Utilizing this tool, we reveal that strong rotation does more than just shift the plasma; it compresses the flux surfaces in a way that forces the core safety factor $q_0$ to drop dangerously close to unity, a phenomenon with clear implications for sawtooth stability.

The remainder of this paper details the methodology and its applications. Section 2 outlines the physical model. Section 3 describes the spectral formulation and matrix acceleration; Section 4 provides rigorous benchmarking against finite difference solutions; and Section 5 explores the physics of strongly rotating equilibria through extensive parameter scans.

\section{Physical Model}

We focus on axisymmetric tokamak plasmas undergoing strong toroidal rotation. Within the framework of ideal magnetohydrodynamics (MHD), assuming the fluid rotates rigidly in the toroidal direction, the equilibrium state is governed by the Generalized Grad-Shafranov (GGS) equation\cite{Maschke1980,Hameiri1983}, which accounts for inertial forces.

\subsection{The Generalized Grad-Shafranov Equation}
Following the magnetohydrodynamic (MHD) formulation described by Jardin\cite{Jardin2010}, we adopt a cylindrical coordinate system $(R,\phi,Z)$. The axisymmetric magnetic field is expressed as $\mathbf{B}=\nabla\phi\times\nabla\psi+F(\psi)\nabla\phi$, where $\psi$ is the poloidal flux and $F(\psi)=RB_\phi$ is the poloidal current function.

Starting from the steady-state momentum equation $\rho\mathbf{v}\cdot\nabla\mathbf{v}=\mathbf{J}\times\mathbf{B}-\nabla p$, we derive the elliptic partial differential equation for the flux distribution\cite{Maschke1980}:
\begin{equation}
\Delta^*\psi = \mu_0 R J_\phi
\end{equation}
Here, the Grad-Shafranov operator is defined as
\begin{equation}
\Delta^*\psi \equiv R \frac{\partial}{\partial R}\left(\frac{1}{R}\frac{\partial\psi}{\partial R}\right) + \frac{\partial^2\psi}{\partial Z^2}
\end{equation}
For a plasma with toroidal flow $\mathbf{v}=R\Omega(\psi)\hat{\phi}$, the expression for the toroidal current density $J_\phi$ requires a crucial modification:
\begin{equation}
J_\phi = - R \left.\frac{\partial P}{\partial\psi}\right|_R - \frac{F(\psi)}{\mu_0 R}\frac{dF}{d\psi}
\end{equation}
Note that the pressure $P(R,\psi)$ is no longer a flux function alone; it explicitly depends on the major radius $R$. This arises because the centrifugal force from high-speed rotation drives a pressure gradient along the major radius, breaking the force balance along magnetic field lines.

\subsection{Thermodynamic Profiles and Rotation Effects}
To close the system, we need an equation of state. Assuming the plasma temperature remains constant on magnetic surfaces, $T=T(\psi)$, the force balance along field lines yields a Bernoulli-type expression for the total pressure. This form is derived by Maschke and Perrin\cite{Maschke1980} and is consistent with the centrifugal density distribution described in neoclassical transport theory by Hinton and Wong\cite{Hinton1985}:
\begin{equation}
P(R,\psi) = P_0(\psi) \exp\left( \frac{M^2(\psi)}{2} \left( \frac{R^2}{R_0^2} - 1 \right) \right)
\end{equation}
In this expression:
\begin{itemize}
    \item $P_0(\psi)$ represents the reference pressure profile, typically corresponding to the static case.
    \item $M(\psi)$ is the toroidal Mach number, defined as the ratio of the fluid velocity to the sound speed $c_s$:
    \begin{equation}
    M(\psi) \equiv \frac{v_\phi}{c_s} = \frac{\Omega(\psi)R_0}{\sqrt{k_B T(\psi)/m_i}}
    \end{equation}
    \item $R_0$ denotes the reference major radius. While standard literature often defines $R_0$ generically as a ``suitably chosen scale length''\cite{Maschke1980}, in this work, we explicitly define $R_0$ as the geometric center of the vacuum chamber. This specific choice serves as the normalization scale for our spectral expansion and ensures consistency across the dimensionless variable definitions.
    \item $m_i$ is the ion mass (in kg), $T(\psi)$ is the temperature (in K), and $k_B$ is the Boltzmann constant (in J/K).
\end{itemize}

The exponential term is the physical engine behind the strong rotation effects. As analyzed by Waelbroeck\cite{Waelbroeck1996}, when the Mach number approaches unity (sonic flow), this term forces the pressure to peak significantly on the low-field side (LFS). This ``centrifugal shift'' of the pressure profile generates the extra torque that drives the non-linear Shafranov shift and geometric distortion observed in our results.

\subsection{Input Decoupling for Thermodynamic Consistency}
We designed the solver's input strategy to align with experimental data analysis and transport modeling needs. Rather than specifying $P(\psi)$ and $M(\psi)$ directly—which can lead to inconsistencies—our model accepts independent inputs for the reference pressure $P_0(\psi)$, ion temperature $T(\psi)$, and toroidal angular velocity $\Omega(\psi)$. The solver then self-consistently calculates the squared Mach number profile $M^2(\psi)$. This approach ensures physical self-consistency, guaranteeing that the derived Mach number profile $M(\psi)$ perfectly matches the specified ion temperature $T(\psi)$ and angular velocity $\Omega(\psi)$ distributions. Numerically, we fit the calculated $M^2(\psi)$ with high-order polynomials to ensure smooth analytic derivatives for the Jacobian matrix calculations.

\section{Numerical Implementation: High-Order Spectral Moments}

Following the theory framework by Xie and Li\cite{Xie2026}, we employ a Shifted Chebyshev Spectral Expansion coupled with a Matrix-Kernel acceleration scheme to solve the non-linear GGS equation on a millisecond timescale while retaining the geometric flexibility to describe strong rotation.

\subsection{Inverse Coordinate Formulation}
Under conditions of strong rotation or high beta, flux surfaces exhibit complex non-linear distortions. While conventional equilibrium codes like EFIT\cite{Lao1985} typically rely on Picard iteration over fixed spatial grids, we adopt an inverse coordinate representation to explicitly capture these geometric deformations. We parameterize the geometry mapping the computational domain $(\rho,\theta)$ to the physical domain $(R,Z)$ via a generalized angle. The radial profiles are described as:
\begin{align}
R(\rho,\theta) &= R_0 + h(\rho) + a\rho \cos \bar{\theta} \\
Z(\rho,\theta) &= Z_0 + v(\rho) - a\kappa(\rho)\rho \sin \bar{\theta}
\end{align}
Here, the generalized angle $\bar{\theta}$ includes a series of angular corrections to describe triangularity and higher-order harmonics:
\begin{equation}
\bar{\theta} = \theta + c_0(\rho) + c_1(\rho)\cos\theta + s_1(\rho)\sin\theta + \dots
\end{equation}
Unlike the traditional variational model proposed by Lao et al.\cite{Lao1981} and Haney et al.\cite{Haney1989,Haney1995}, this parameterization is more general, allowing us to describe asymmetric flux surfaces with higher degrees of freedom, defined uniquely by radial functions such as the Shafranov shift $h(\rho)$, elongation $\kappa(\rho)$, and shaping coefficients $s_n(\rho)$.

To rigorously define the equilibrium problem, we define the coordinate transformation Jacobian $J$. Consistent with the right-handed system $(R,\phi,Z)$, the Jacobian is explicitly defined as:
\begin{equation}
J = \frac{\partial R}{\partial \theta}\frac{\partial Z}{\partial \rho} - \frac{\partial R}{\partial \rho}\frac{\partial Z}{\partial \theta}
\end{equation}
This Jacobian ensures the conservation of volume elements via $dV = 2\pi R J d\rho d\theta$. Using the chain rule, the spatial differential operators in the physical frame are transformed into the computational frame as follows:
\begin{equation}
\frac{\partial}{\partial R} = \frac{1}{J}\left( \frac{\partial Z}{\partial \rho}\frac{\partial}{\partial \theta} - \frac{\partial Z}{\partial \theta}\frac{\partial}{\partial \rho} \right), \quad \frac{\partial}{\partial Z} = \frac{1}{J}\left( \frac{\partial R}{\partial \theta}\frac{\partial}{\partial \rho} - \frac{\partial R}{\partial \rho}\frac{\partial}{\partial \theta} \right)
\end{equation}
Applying these operators to the Grad-Shafranov operator $\Delta^*\psi = R^2\nabla\cdot(R^{-2}\nabla\psi)$ and imposing the flux coordinate assumption ($\psi=\psi(\rho)$), we derive the Generalized Grad-Shafranov (GGS) equation in inverse coordinates:
\begin{equation}
\frac{R}{J}\frac{\partial}{\partial\rho}\left( \frac{R_\theta^2+Z_\theta^2}{JR}\psi_\rho \right) - \frac{\partial}{\partial\theta}\left( \frac{R_\rho R_\theta+Z_\rho Z_\theta}{JR}\psi_\rho \right) = \mu_0 R J_\phi
\end{equation}
The toroidal current density $J_\phi$ on the right-hand side is evaluated locally at each computational node $(\rho,\theta)$ and explicitly incorporates the centrifugal effects. Its expanded form is:
\begin{equation}
J_\phi(\rho,\theta) = -R\frac{dP_0}{d\psi}E_{rot} - P_0(\psi)E_{rot}\frac{d}{d\psi}\left[ \frac{M^2}{2}\left( \frac{R^2}{R_0^2}-1 \right) \right] - \frac{1}{\mu_0 R}F\frac{dF}{d\psi}
\end{equation}
where $E_{rot}$ denotes the centrifugal enhancement factor defined as:
\begin{equation}
E_{rot}(\rho,\theta) = \exp\left( \frac{M^2(\psi)}{2}\left( \frac{R(\rho,\theta)^2}{R_0^2}-1 \right) \right)
\end{equation}
This formulation highlights the strong non-linearity of the problem: the current density driving the equilibrium is itself dependent on the geometric solution $R(\rho,\theta)$ through the centrifugal potential term.

\subsection{Global Spectral Expansion via Chebyshev Polynomials}
For high-precision spectral approximation, we expand both the geometric functions and the poloidal flux $\psi(\rho)$ using Shifted Chebyshev Polynomials, following the spectral methods outlined by Boyd\cite{Boyd2001}. An arbitrary profile $f(\rho)$ is expressed as:
\begin{equation}
f(\rho) = \rho^p \left( f_a + \sum_{l=0}^{L} f_l u_l(\rho) \right)
\end{equation}
where $p$ is a parity factor ensuring analyticity at the axis, and $f_a$ represents the boundary or axis value. The basis functions $u_l(\rho)$ are constructed as the product of Chebyshev polynomials $T_l(x)$ and a boundary constraint factor $(1-\rho^2)$:
\begin{equation}
u_l(\rho) = (1-\rho^2) \cdot T_l(2\rho^2-1)
\end{equation}
Here, $T_l(x)$ denotes the Chebyshev polynomials of the first kind ($T_0=1, T_1=x, T_2=2x^2-1, \dots$). This specific construction exploits the optimal uniform approximation property of Chebyshev polynomials on the interval $[-1,1]$\cite{Boyd2001,Canuto2006}, which is instrumental in significantly accelerating the convergence rate of the algorithm.

In the proposed 12-parameter model, the specific forms of these expansions are tailored to naturally enforce geometric regularity and physical boundary conditions.
For the Shafranov shift $h(\rho)$, we employ an even-parity expansion where the boundary condition $h(1)=0$ is intrinsically satisfied by the basis definition:
\begin{equation}
h(\rho) = (1-\rho^2)\left( h_0 + h_1(2\rho^2-1) + \dots \right)
\end{equation}
Simultaneously, for the elongation $\kappa(\rho)$, we allow the profile to vary radially from the fixed boundary value $\kappa_a$. This freedom is essential for capturing the dynamic stretching of the core plasma induced by rotation, represented as
\begin{equation}
\kappa(\rho) = \kappa_a + (1-\rho^2)\left( \kappa_0 + \kappa_1(2\rho^2-1) + \dots \right)
\end{equation}
To describe higher-order deformations such as triangularity $s_1(\rho)$ while maintaining geometric regularity at the magnetic axis (i.e., non-singularity as $\rho\to0$), a factor of $\rho$ is explicitly introduced into the expansion:
\begin{equation}
s_1(\rho) = \rho\left[ s_{1a} + (1-\rho^2)\left( s_{10} + s_{11}(2\rho^2-1) + \dots \right) \right]
\end{equation}
where $s_{1a}$ corresponds to the triangularity at the plasma boundary. Finally, the poloidal flux $\psi(\rho)$ is expanded to automatically satisfy the zero-shear condition at the axis ($\psi'(0)=0$) and match the boundary flux $\psi_a$, ensuring the solution respects the asymptotic behavior required by the GGS equation:
\begin{equation}
\psi(\rho) = \psi_a \rho^2 \left[ 1 + (1-\rho^2)\left( v_0 + v_1(2\rho^2-1) + \dots \right) \right]
\end{equation}
By retaining high-order terms (e.g., up to $l=1$ or $l=2$), this formulation constructs a compact parameter space of 12 to 15 degrees of freedom. The inclusion of second-order Chebyshev terms like $T_2(2\rho^2-1)$ provides the necessary mathematical foundation to resolve the non-rigid geometric distortions excited under strong centrifugal loads.

\subsection{Weighted Residual Formulation and Generalized Galerkin Projection}
Solving the equilibrium problem is formally equivalent to finding the root of the Generalized Grad-Shafranov residual within the discretized parameter space. This approach builds upon the rigorous variational foundation established by Lao et al.\cite{Lao1981} and Haney et al.\cite{Haney1989,Haney1995}, where the equilibrium solution implies the minimization of the system's energy functional. In our unified solver, we adopt a Generalized Galerkin Method that treats both the magnetic flux and the geometric boundary coefficients as coupled variables in a single solution vector $\mathbf{x}$. We define the scalar residual field $\mathcal{G}$ over the computational domain $(\rho,\theta)$ as the local error of the force balance equation:
\begin{equation}
\mathcal{G} \equiv \frac{R}{J}\frac{\partial}{\partial\rho}\left(\frac{R_\theta^2+Z_\theta^2}{JR}\psi_\rho\right) - \frac{\partial}{\partial\theta}\left(\frac{R_\rho R_\theta+Z_\rho Z_\theta}{JR}\psi_\rho\right) + FF'(\psi) + R^2\mu_0 P'(\psi)
\end{equation}
where the source term components are consistent with the thermodynamic model described in Section 2. To determine the spectral coefficients, we impose the Galerkin condition that the projection of this residual onto the parameter sensitivity basis must vanish. Specifically, the weight function $w_k$ corresponding to each unknown parameter $x_k$ is derived as the total derivative of the poloidal flux with respect to that parameter, i.e., $w_k(\rho,\theta) = d\psi(R(\mathbf{x}),Z(\mathbf{x}); \mathbf{x})/dx_k$. This projection yields the algebraic system:
\begin{equation}
\mathcal{R}_k = \int_0^1 d\rho \int_0^{2\pi} d\theta \, J \cdot \mathcal{G} \cdot w_k = 0
\end{equation}
By applying the chain rule, this formulation naturally recovers the specific Galerkin projection forms for the two classes of coefficients. For the flux spectral coefficients $c_\psi$, since $\psi$ depends linearly on its expansion, the weight function is simply the spectral basis function itself ($w_{c_\psi} = \phi_k(\rho)$), corresponding to the standard Galerkin projection for the magnetic flux equation. For the geometric coefficients $c_{geo}$ (such as those defining the Shafranov shift $h$ or elongation $\kappa$), the weight function becomes the convective derivative $w_{c_{geo}} = \nabla\psi \cdot \partial\mathbf{r}/\partial c_{geo}$. In our ``Matrix-Kernel'' numerical implementation, the term $J \cdot \mathcal{G}$ represents the weighted residual density, which is then projected onto these transformed basis vectors. This Generalized Galerkin formulation ensures that the geometry dynamically evolves to minimize the force balance error along the direction of the magnetic field gradient, providing robust convergence even for highly distorted plasmas in the presence of strong rotation.

\subsection{Matrix-Kernel Acceleration Scheme}
To solve these coupled integral equations efficiently without the computational bottleneck of numerical quadrature, we introduce a Matrix-Kernel algorithm. By exploiting the orthogonality of the basic functions and the static nature of the discretization grid, we act to fully ``matrixize'' the integration process. We first discretize the computational domain onto a pre-calculated Gaussian integration grid, where the GGS source error (including the Laplacian $\Delta^*$ and centrifugal terms $J_\phi$) is evaluated as a generalized residual vector $\mathbf{G}_{grid}$. The residual evaluation is then decomposed into two parallel, highly vectorized matrix operation channels using pre-computed projection matrices $\mathbf{M}$. The flux residual is obtained by directly projecting the weighted grid error,
\begin{equation}
\mathcal{R}_\psi = \mathbf{M}_\psi^T \cdot (\mathbf{W} \odot \mathbf{G}_{grid})
\end{equation}
Simultaneously, the geometric residual incorporates the local flux gradient field to account for the coordinate variations, calculated as
\begin{equation}
\mathcal{R}_{geo} = \mathbf{M}_{geo}^T \cdot (\mathbf{W} \odot \mathbf{G}_{grid} \odot \nabla\psi \cdot J_{coord})
\end{equation}
where $\mathbf{W}$ is the weight vector, $\odot$ denotes the Hadamard product, and $J_{coord}$ is the coordinate transformation Jacobian. This formulation transforms the traditionally expensive variational integration into efficient algebraic matrix multiplications, effectively eliminating loop operations and reducing the single-pass residual evaluation time to the microsecond scale.

\subsection{Robust Solver Strategy}
The inclusion of high-order non-linear terms and strong rotation effects significantly increases the stiffness of the equation system. To ensure convergence, we employ a hybrid solver strategy based on Broyden’s quasi-Newton method\cite{Broyden1965} and advanced iterative schemes\cite{Kelley1995}. A critical challenge in high-rotation regimes ($M_{axis}>1.0$) is the potential near-singularity of the Jacobian matrix, which arises from the extreme sensitivity of the equilibrium to centrifugal distortions. To mitigate this, we integrate Truncated SVD Regularization\cite{Hansen1998} into the solver. This technique filters out the ill-conditioned modes associated with numerical noise while preserving the dominant physical modes, thereby significantly enhancing the algorithm's robustness and ensuring stable convergence even under conditions of sonic rotation.

\section{Benchmarking and Validation}

To rigorously evaluate the accuracy and reliability of the proposed 12-parameter VEQ-R solver, we implemented a hierarchical validation strategy. Since the generalized Grad-Shafranov equation with strong centrifugal non-linearity does not admit analytic solutions, we developed a high-resolution Finite Difference Method (FDM) solver to serve as the numerical ``Ground Truth'' for validating the new algorithm across different rotation regimes.

\subsection{Reference Solver: High-Fidelity FDM}
To eliminate uncertainties arising from model discrepancies, the benchmark solver was designed to govern the exact same physical equations described in Section 2. We discretized the computational domain onto a high-resolution rectangular grid ($513 \times 513$), approximating the Grad-Shafranov operator via a standard five-point finite difference stencil with a second-order truncation error of $O(\Delta h^2)$, which transforms the partial differential equation into a large-scale linear system with a highly sparse connectivity matrix.

For the non-linear source term $J_\phi(\psi,R)$, we employed a Picard iteration scheme. At each iteration step, the resulting sparse linear system is solved using a direct multifrontal method, a technique optimized for such sparse structures as detailed by Davis\cite{Davis2006}. This rigorous approach ensures a robust convergence with a strict criterion of $\|\psi^{k+1}-\psi^k\|_\infty < 10^{-10}$.

To ensure a rigorous benchmark, both the VEQ-R and FDM solvers are initialized with identical geometric parameters and physical profile inputs. As illustrated in Figure 1, we enforce thermodynamic consistency by deriving the Mach number profile from the self-consistent temperature $T(\psi)$ and rotational angular velocity $\Omega(\psi)$ distributions, ensuring that the centrifugal effects are consistently modeled in both solvers.

\begin{figure}[H]
    \centering
    \includegraphics[width=0.85\linewidth, height=0.3\textheight, keepaspectratio]{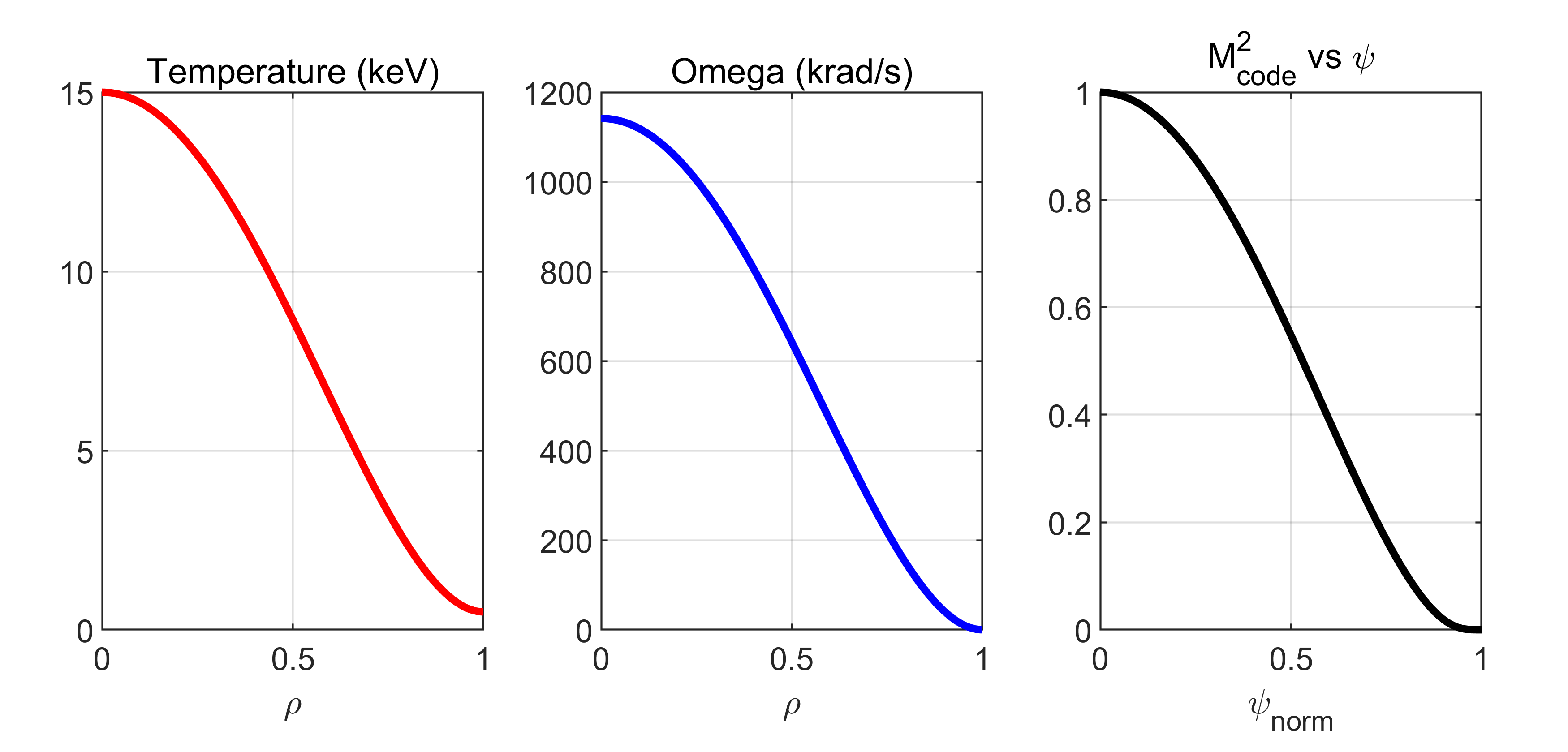}
    \caption{Input physical profiles used for benchmarking the solver. (a) Ion temperature profile $T(\rho)$; (b) Toroidal angular velocity profile $\Omega(\rho)$; (c) The self-consistently calculated squared Mach number profile $M^2(\psi)$ derived from inputs (a) and (b). Note that the initialization ensures thermodynamic consistency where the Mach number peaks at the magnetic axis.}
\end{figure}

To facilitate reproducibility, we explicitly define the boundary conditions and profile parameterizations used in this study. The geometric boundary is parameterized by the standard shaping form:
\begin{equation}
R_b = R_0 + a\cos(\theta + \sin^{-1}\delta\sin\theta), \quad Z_b = -\kappa a \sin\theta
\end{equation}
Regarding the internal profiles, we employ two distinct functional forms. The static reference pressure $P_0'(\psi)$ and the poloidal current function term $FF'(\psi)$ utilize a modified exponential parameterization. This form is specifically chosen to allow independent control over core peaking while ensuring smooth gradients at the plasma edge:
\begin{equation}
X(\psi) \propto \alpha_X \frac{e^{\alpha_X\psi}-e^{\alpha_X}}{1+e^{\alpha_X}(\alpha_X-1)}
\end{equation}
where $X$ represents either $P_0'$ or $FF'$, and $\alpha_X$ is the peaking parameter. Let $\psi$ denote the normalized poloidal flux, defined as $\psi = (\psi-\psi_{axis})/(\psi_{edge}-\psi_{axis})$. The profiles for the temperature $T(\psi)$, and toroidal angular velocity $\Omega(\psi)$ are prescribed using the generalized power-law form:
\begin{equation}
Y(\psi) = Y_0(1-\psi^{\alpha_Y})^{\beta_Y} + Y_{edge}
\end{equation}
where $Y$ represents $T$ or $\Omega$. The specific numerical coefficients for all geometric and physical parameters are listed in Table I.

\begin{table}[H]
\centering
\caption{Geometric parameters and physical profile definitions. The pressure and current profiles use the exponential form defined in the text, while temperature and rotation follow a power-law distribution.}
\begin{tabular}{llc}
\toprule
\textbf{Category} & \textbf{Parameter / Symbol} & \textbf{Value} \\
\midrule
Geometry & Major Radius ($R_0$) & 1.05 m \\
         & Minor Radius ($a$) & 0.57 m \\
         & Aspect Ratio ($A$) & 1.85 \\
         & Boundary Elongation ($\kappa_{edge}$) & 2.2 \\
         & Boundary Triangularity ($\delta_{edge}$) & 0.5 \\
\midrule
Physics  & Vacuum Toroidal Field ($B_0$) & 3.0 T \\
         & Plasma Current ($I_p$) & 3.0 MA \\
         & Core Mach Number ($M_0$) & 1.0 (for rotating case) \\
\midrule
Profiles & Pressure Peaking ($\alpha_P$) & 5.0 \\
         & Current Peaking ($\alpha_F$) & 3.32 \\
         & Temperature Index ($\alpha_T, \beta_T$) & 2.0, 2.0 \\
         & Rotation Index ($\alpha_\Omega, \beta_\Omega$) & 2.0, 2.0 \\
\bottomrule
\end{tabular}
\end{table}

\subsection{Model Convergence and Static Verification ($M=0$)}
Before assessing global accuracy, we analyzed the evolution of the spectral coefficients to understand the model's physical response mechanism. Figure 2 displays the variation of key spectral coefficients with the Mach number. The results reveal a distinct behavior: low-order geometric modes, such as the primary Shafranov shift coefficient $C_{h1}$ (shown in Figure 2a), exhibit significant non-linear growth driven by the centrifugal force. Crucially, the coefficients representing high-order geometric fine-tuning—such as the high-order elongation $C_{\kappa3}$ and triangularity $C_{\delta3}$ (shown in Figure 2b/c)—maintain non-negligible magnitudes and follow clear monotonic trends throughout the parameter space. This confirms that the 12-parameter model successfully captures the ``non-rigid'' geometric distortions via dynamic adaptation, whereas traditional low-order models\cite{Haney1989,Haney1995} would suffer from systematic bias due to the lack of these degrees of freedom.

\begin{figure}[H]
    \centering
    \includegraphics[width=0.85\linewidth, height=0.3\textheight, keepaspectratio]{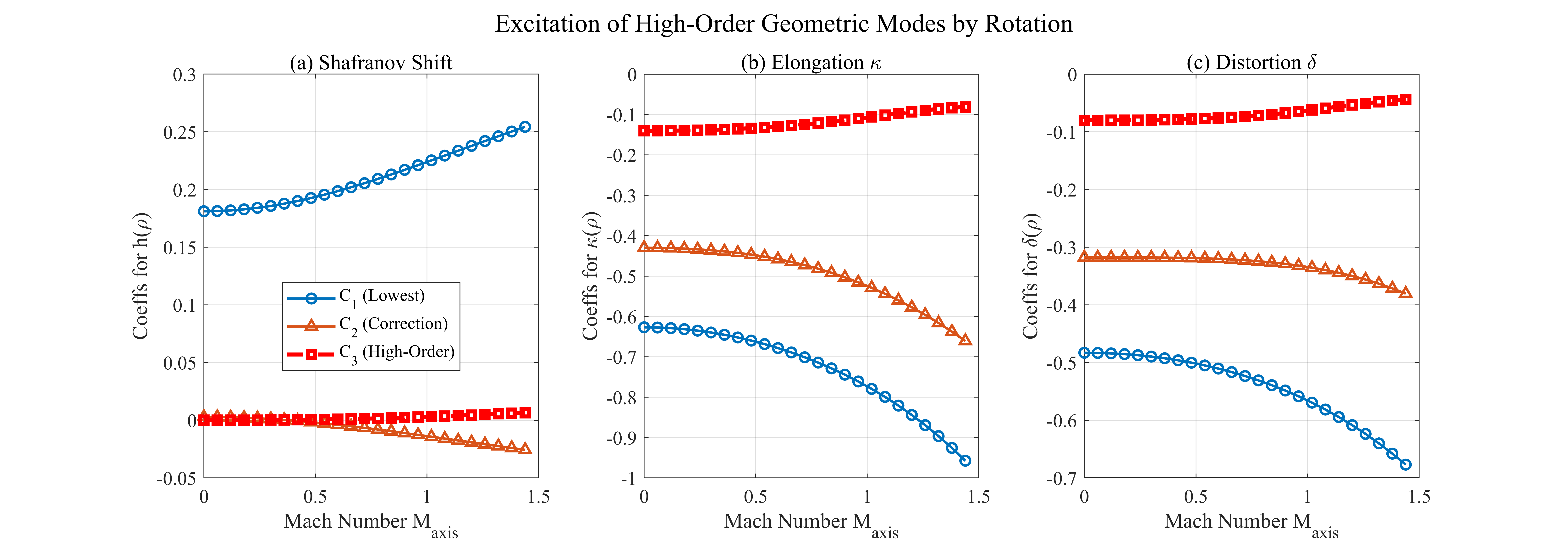}
    \caption{Evolution of spectral geometric coefficients with increasing axis Mach number $M_{axis}$. (a) Shafranov shift coefficients ($h$); (b) Elongation coefficients ($\kappa$); (c) Triangularity/Distortion coefficients ($\delta$). The distinct monotonic trends of the high-order harmonic terms (red lines, e.g., $C_3$) confirm that the 12-parameter VEQ-R model successfully captures the ``non-rigid'' geometric distortions excited by centrifugal forces.}
\end{figure}

Table II lists the converged spectral coefficients for the static baseline ($M_0=0$). In this regime, the geometric deformation is dominated by the lowest-order harmonics ($c_1$), and the high-order correction terms ($c_3$) for the Shafranov shift are negligible, confirming the smoothness of the static flux surfaces.

\begin{table}[H]
\centering
\caption{Converged spectral coefficients for the static baseline case ($M_0=0$).}
\begin{tabular}{llc}
\toprule
\textbf{Profile Function} & \textbf{Basis Order} & \textbf{Coefficient Value ($M_0=0$)} \\
\midrule
Shafranov Shift $h(\rho)$ & $c_1$ (Lowest) & 0.1811 \\
                          & $c_2$ (Correction) & 0.0029 \\
                          & $c_3$ (High-Order) & 0.0000 \\
\midrule
Elongation $\kappa(\rho)$ & $c_1$ (Lowest) & -0.6270 \\
                          & $c_2$ (Correction) & -0.4300 \\
                          & $c_3$ (High-Order) & -0.1403 \\
\midrule
Triangularity $s_1(\rho)$ & $c_1$ (Lowest) & -0.4830 \\
                          & $c_2$ (Correction) & -0.3178 \\
                          & $c_3$ (High-Order) & -0.0800 \\
\midrule
Poloidal Flux $\psi(\rho)$ & $c_1$ (Lowest) & 0.5220 \\
                           & $c_2$ (Correction) & -0.0335 \\
                           & $c_3$ (High-Order) & 0.0101 \\
\bottomrule
\end{tabular}
\end{table}

Having established the spectral convergence, we verified the solver's accuracy under static conditions. Figure 3 presents the comparison of magnetic flux contours, where the VEQ-R reconstruction (blue solid lines) shows excellent agreement with the FDM reference (red dashed lines) across the entire domain. The absolute deviation of the magnetic axis position is merely $9 \times 10^{-4}$ m, confirming that the 12-parameter spectral basis is sufficient to resolve standard D-shaped tokamak geometries.

\begin{figure}[H]
    \centering
    \includegraphics[width=0.85\linewidth, height=0.3\textheight, keepaspectratio]{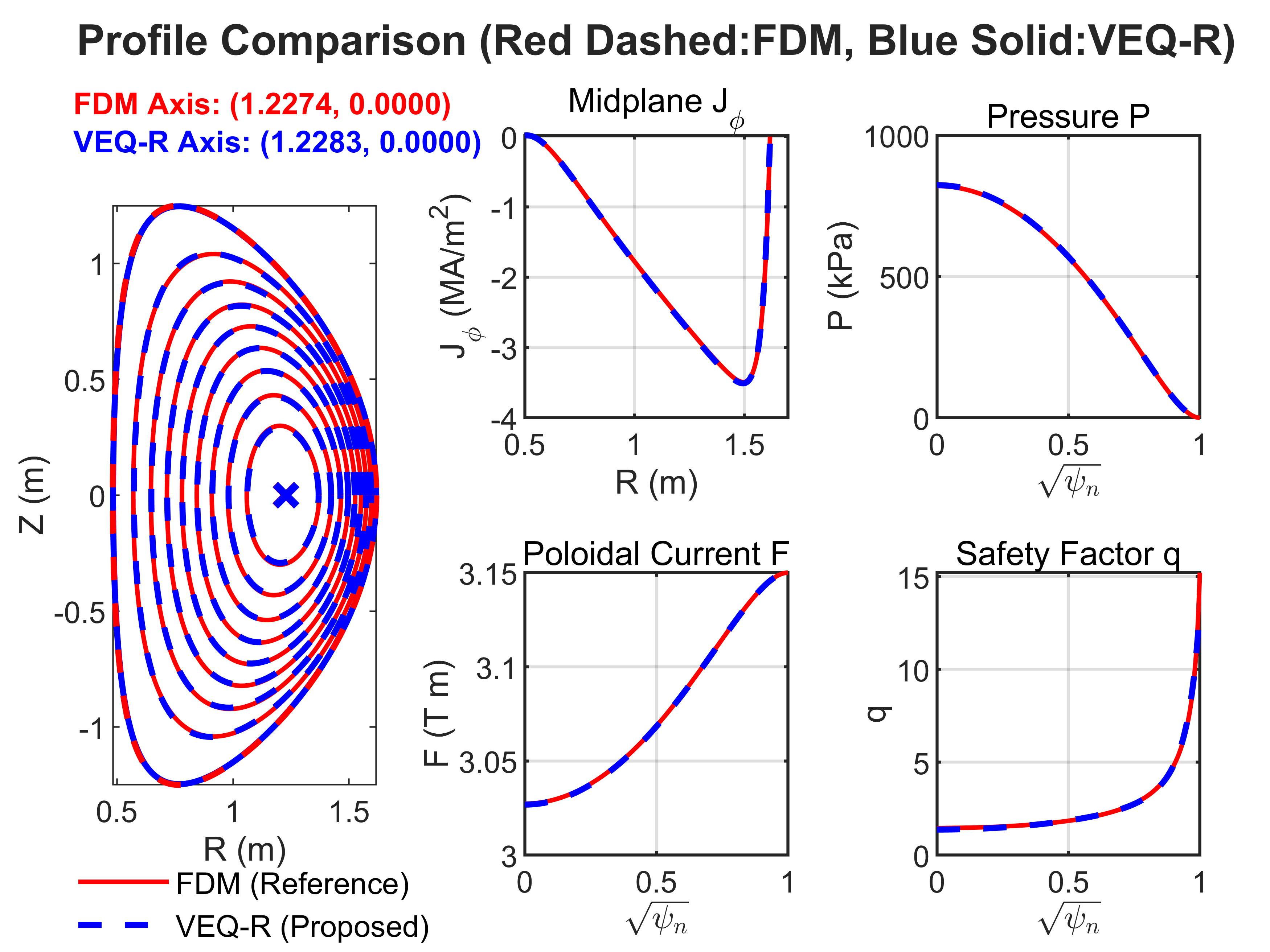}
    \caption{Benchmark results under static conditions ($M=0$). Left: Comparison of magnetic flux surfaces between the proposed VEQ-R solver (blue solid lines) and the high-resolution FDM reference (red dashed lines), showing excellent geometric agreement. Right: Comparison of midplane profiles for toroidal current density $J_\phi$, pressure $P$, poloidal current function $F$, and safety factor $q$.}
\end{figure}

To quantify the precision, Figure 4 analyzes the radial distribution of relative errors for key physical quantities. In the core region ($\rho < 0.9$), the relative errors for pressure $P$, poloidal current function $F$, and safety factor $q$ remain consistently below 1\%. While a localized error increase ($\sim 2-3\%$) is observed near the plasma edge ($\rho > 0.95$) due to the Gibbs phenomenon characteristic of spectral methods, its impact on global integrals such as total stored energy $W_{MHD}$ and plasma current $I_p$ is negligible ($< 0.05\%$), satisfying the requirements for core control applications.

\begin{figure}[H]
    \centering
    \includegraphics[width=0.85\linewidth, height=0.3\textheight, keepaspectratio]{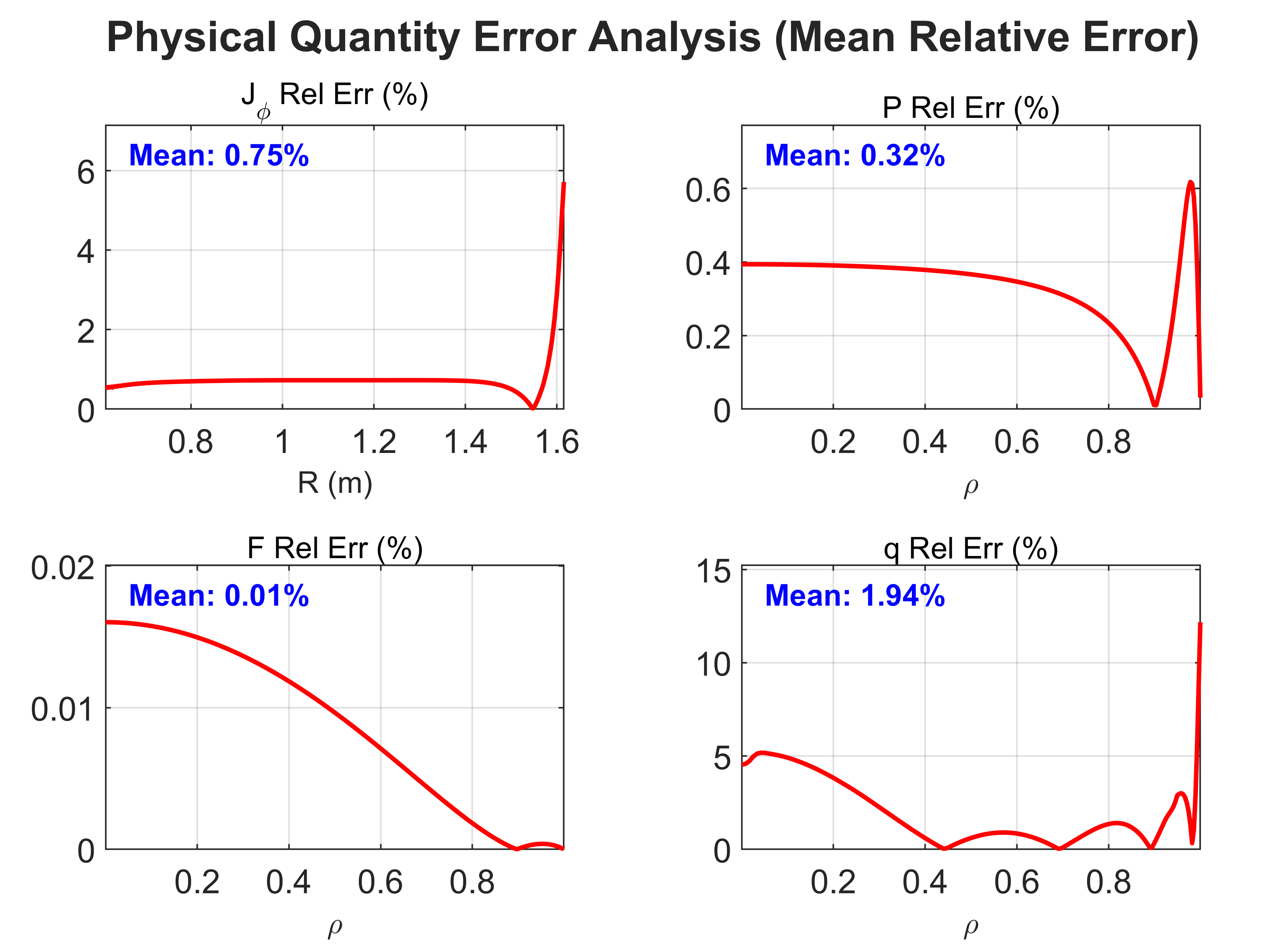}
    \caption{Quantitative error analysis for the static case ($M=0$). The subplots display the radial distribution of relative errors for (top-left) toroidal current density $J_\phi$, (top-right) pressure $P$, (bottom-left) poloidal current function $F$, and (bottom-right) safety factor $q$. The mean relative error for each quantity is indicated, demonstrating high accuracy in the absence of rotation.}
\end{figure}

\subsection{Verification of Rotating Equilibria ($M \approx 1$)}
The converged coefficients for the sonic rotation case ($M_0=1.0$) are presented in Table III. Comparing these with the static baseline, we observe a significant excitation of the high-order shift coefficient ($h_{c3}=0.0030$). This quantitatively captures the non-rigid compression of flux surfaces driven by the centrifugal force, a feature that simplified models often miss.

\begin{table}[H]
\centering
\caption{Converged spectral coefficients for the sonic rotation case ($M_0=1.0$), showing the excitation of higher-order harmonics due to centrifugal effects.}
\begin{tabular}{llc}
\toprule
\textbf{Profile Function} & \textbf{Basis Order} & \textbf{Coefficient Value ($M_0=1.0$)} \\
\midrule
Shafranov Shift $h(\rho)$ & $c_1$ (Lowest) & 0.2239 \\
                          & $c_2$ (Correction) & -0.0137 \\
                          & $c_3$ (High-Order) & 0.0030 \\
\midrule
Elongation $\kappa(\rho)$ & $c_1$ (Lowest) & -0.7732 \\
                          & $c_2$ (Correction) & -0.5244 \\
                          & $c_3$ (High-Order) & -0.1073 \\
\midrule
Triangularity $s_1(\rho)$ & $c_1$ (Lowest) & -0.5657 \\
                          & $c_2$ (Correction) & -0.3342 \\
                          & $c_3$ (High-Order) & -0.0630 \\
\midrule
Poloidal Flux $\psi(\rho)$ & $c_1$ (Lowest) & 0.7084 \\
                           & $c_2$ (Correction) & -0.1458 \\
                           & $c_3$ (High-Order) & 0.0257 \\
\bottomrule
\end{tabular}
\end{table}

To stress-test the solver, we simulated a transonic rotation scenario with a core squared Mach number of $M^2=1.0$. This extreme condition imposes dual challenges: severe geometric distortion and thermodynamic non-equilibrium. Figure 5 presents the comparison for this high-rotation case. Under the strong centrifugal force, the magnetic axis shifts significantly toward the low-field side. The FDM reference places the axis at $R=1.2872$ m, while the VEQ-R calculates it at $1.2906$ m. The discrepancy is limited to 3.4 mm (approximately 0.8\% of the minor radius), demonstrating that the algorithm accurately captures the non-linear Shafranov shift. Simultaneously, the VEQ-R accurately reproduces the pile-up characteristics of pressure $P$ and toroidal current density $J_\phi$ on the low-field side.

\begin{figure}[H]
    \centering
    \includegraphics[width=0.85\linewidth, height=0.3\textheight, keepaspectratio]{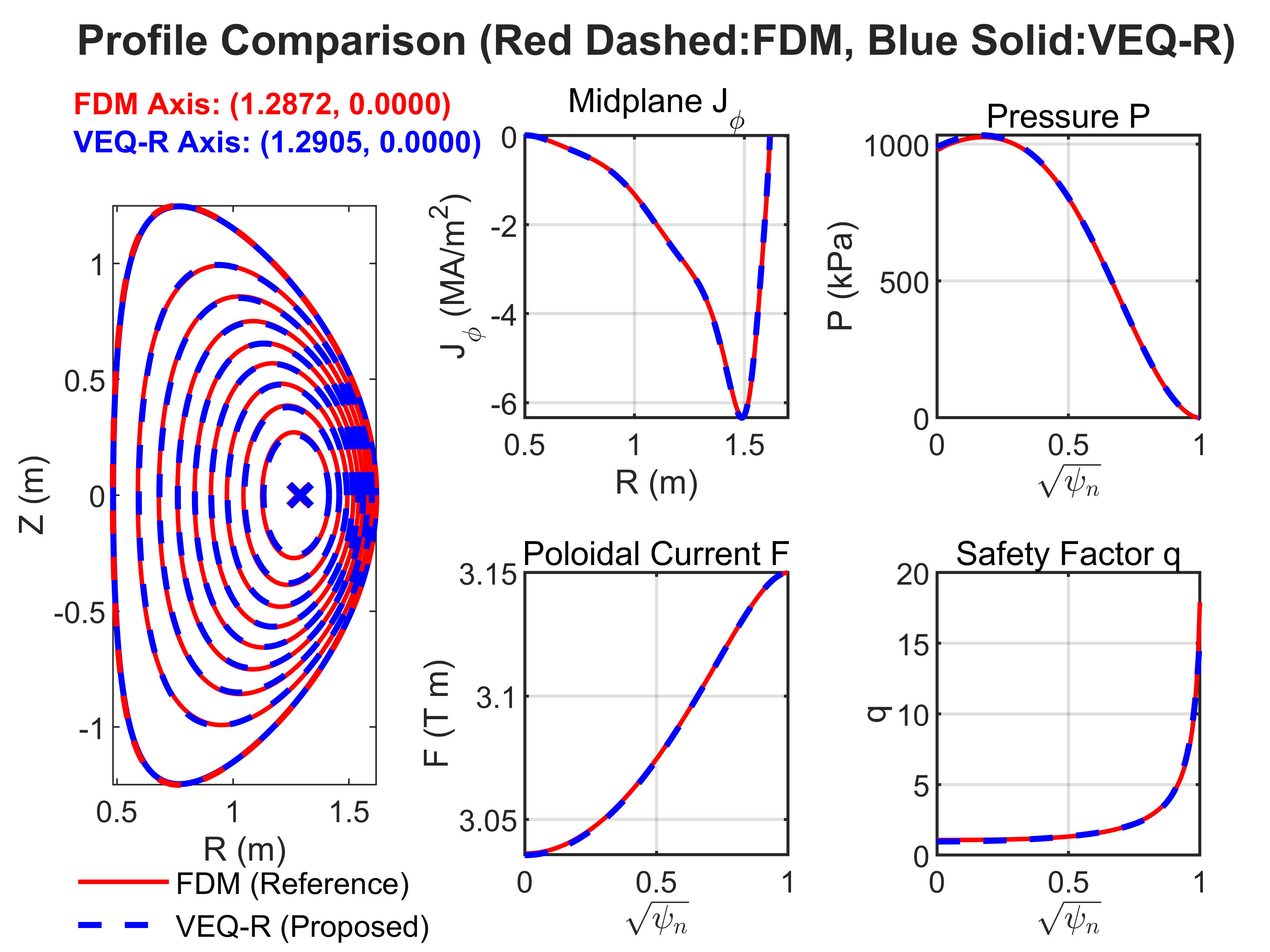}
    \caption{Benchmark results under transonic rotation conditions ($M_{axis}^2=1.0$). Left: Flux surface comparison showing the significant outward Shafranov shift and geometric compression on the low-field side driven by centrifugal forces. Right: Comparison of key physical profiles ($J_\phi, P, F, q$), illustrating the VEQ-R solver's capability to accurately reconstruct the asymmetric current pile-up and pressure shifts.}
\end{figure}

The corresponding quantitative error analysis is detailed in Figure 6. The solver maintains high fidelity even in this highly non-linear regime. The average relative errors for integral quantities like the poloidal current function $F$ are as low as 0.01\%, and the pressure $P$ error is only 0.37\%, indicating precise satisfaction of flux conservation and force balance. The toroidal current $J_\phi$ error is controlled at 0.84\%. Although the safety factor $q$ exhibits a slightly higher average error of 2.85\%—concentrated mainly near the magnetic axis and the edge due to sensitivity to flux gradients—the global shape and magnitude of the $q$-profile are accurately reconstructed, meeting the fidelity requirements for transport simulations and stability analysis.

\begin{figure}[H]
    \centering
    \includegraphics[width=0.85\linewidth, height=0.3\textheight, keepaspectratio]{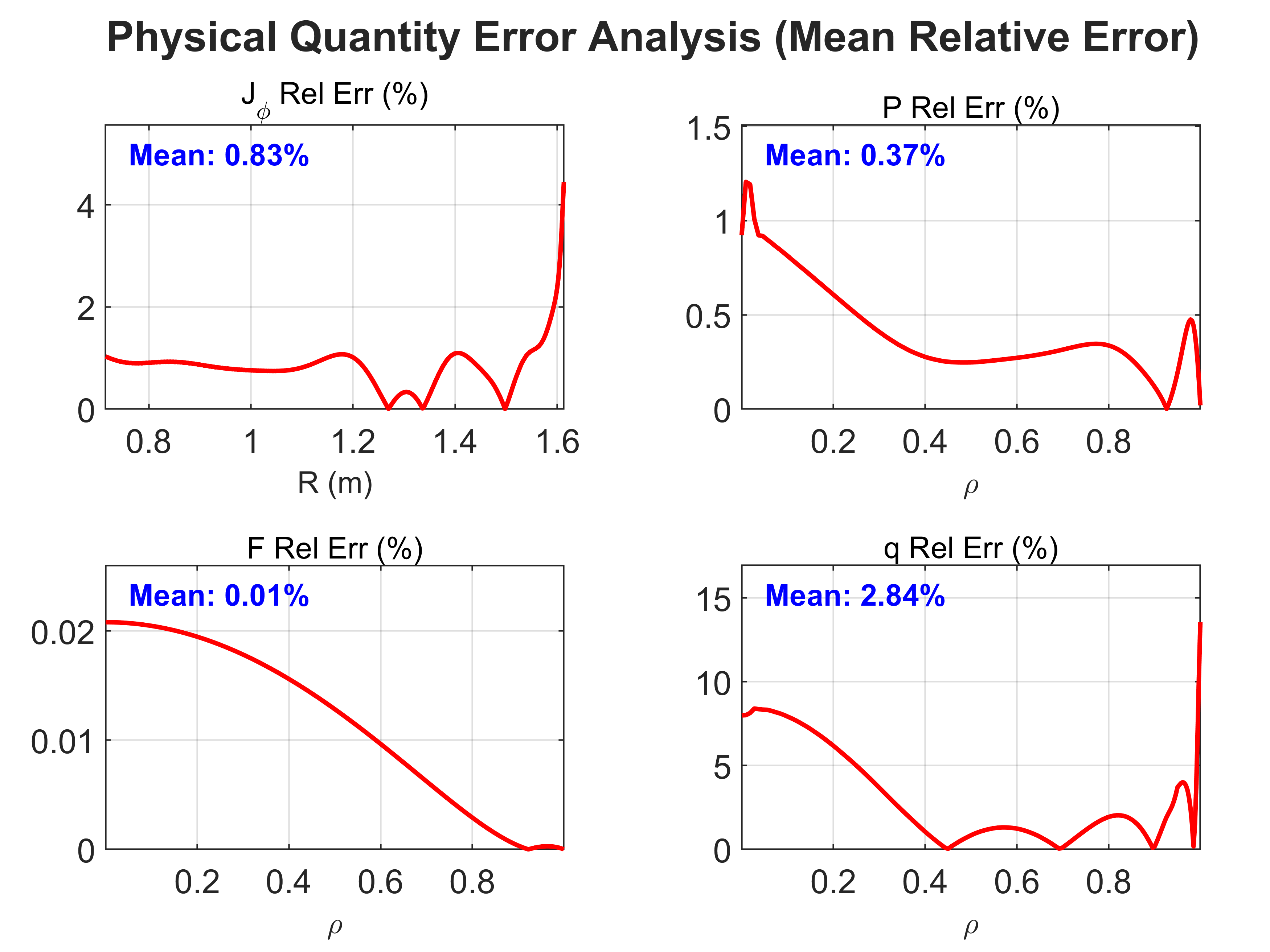}
    \caption{Quantitative error analysis for the transonic rotation case ($M_{axis}^2=1.0$). Despite the severe geometric distortion, the VEQ-R solver maintains low relative errors across all physical profiles. The safety factor $q$ exhibits a slightly higher error near the axis due to vanishing gradients but remains globally well-resolved for stability analysis.}
\end{figure}

\subsection{Computational Efficiency and Comparison}
Finally, we evaluate the performance of the VEQ-R solver in the context of real-time control. Table \ref{tab:runtime} compares the typical runtimes and characteristics of various equilibrium solvers.

\begin{table}[H]
\centering
\caption{Comparison of computational performance, numerical methodology, and typical runtimes between the proposed Matrix-Kernel VEQ-R solver and existing high-fidelity equilibrium codes capable of modeling toroidal flow.}
\label{tab:runtime}
\begin{tabular}{lllc}
\toprule
\textbf{Solver / Reference} & \textbf{Methodology} & \textbf{Grid / DOF} & \textbf{Runtime} \\
\midrule
This Work (VEQ-R) & Spectral + Matrix-Kernel & 12 Parameters & $\sim$ 5 ms \\
Feng et al.\cite{Feng2024} & Inverse Coordinate FDM & High Res ($N \times N$) & $\sim$ 1-200 s \\
FLOW / FINESSE\cite{Guazzotto2004,Belien2002} & Finite Element (FEM) & Adaptive Grid & $\sim$ 1-200 s \\
Varadarajan et al.\cite{Varadarajan1991} & Variational & Low Order & $\sim$ 15 s \\
\bottomrule
\end{tabular}
\end{table}

Compared to high-fidelity inverse coordinate solvers (e.g., Feng et al., 2024\cite{Feng2024}) or standard finite element codes like FLOW\cite{Guazzotto2004} and FINESSE\cite{Belien2002}, which typically require seconds to minutes due to grid generation and iterative overhead, our Matrix-Kernel VEQ-R solver achieves a convergence time of approximately 5 ms by MATLAB, and further speed up to sub-ms is possible using C++ or Fortran. This represents a speedup factor of roughly 1000. Furthermore, unlike early variational models (e.g., Varadarajan et al., 1991\cite{Varadarajan1991}) that relied on slow steepest descent optimization ($\sim 15$ s), our approach utilizes a Broyden’s quasi-Newton method on the spectral coefficients, achieving a qualitative leap in speed. This millisecond-level capability confirms the potential of the VEQ-R algorithm for deployment in the Plasma Control Systems (PCS) of next-generation devices.

\section{Results and Discussion}

Having validated the VEQ-R solver against high-fidelity benchmarks, we now leverage its computational efficiency to perform a comprehensive parameter scan. This section explores the non-linear reshaping of the plasma equilibrium by toroidal rotation, extending the analysis from static conditions ($M_{axis}=0$) into the supersonic regime ($M_{axis}=1.4$). Our objective is to disentangle the complex interplay between macroscopic geometric distortion and the microscopic modification of thermodynamic and magnetic profiles under extreme centrifugal loads.

\subsection{Global Equilibrium Reshaping and Stability Response}
The centrifugal force exerts a profound global influence on the plasma configuration, reshaping key macroscopic parameters in a highly non-linear manner as summarized in Fig. 7. The most immediate geometric response is observed in the normalized Shafranov shift, $\Delta R / a$ (Fig. 7a), where $a$ denotes the minor radius. As the rotation accelerates, the centrifugal term $\rho\Omega^2R$ increasingly dominates the radial force balance, driving the plasma column significantly toward the low-field side (LFS). Upon reaching the supersonic regime, this displacement exhibits a supra-linear growth, indicating that the plasma is undergoing severe geometric reconfiguration.

Concurrently, the thermodynamic symmetry is fundamentally broken; the poloidal asymmetry factor ($P_{LFS}/P_{HFS}$), shown in Fig. 7b, rises exponentially, confirming that the pressure distribution on magnetic surfaces has deviated entirely from the static equilibrium assumption. This geometric and thermodynamic restructuring imposes critical constraints on the macroscopic stability of the plasma. A particularly concerning trend is the monotonic degradation of the core safety factor, $q_0$ (Fig. 7c). Driven by the intense compression of flux surfaces against the LFS boundary, the core toroidal current density is forced to concentrate to satisfy flux conservation. Consequently, $q_0$ falls from a robust value of approximately 1.4 in the static case toward unity in high-rotation regimes, signaling a heightened susceptibility to $m/n=1/1$ internal kink modes.

Furthermore, despite the substantial elevation of local pressure peaks, the global normalized beta $\beta_N$ (Fig. 7d) exhibits a slight decline. This reveals a subtle trade-off mechanism: the energetic cost required to maintain equilibrium under such severe geometric distortion outweighs the local pressure gains, suggesting that rotation-induced confinement improvement is limited by global magnetohydrodynamic stability boundaries.

\begin{figure}[H]
    \centering
    \includegraphics[width=0.85\linewidth, height=0.3\textheight, keepaspectratio]{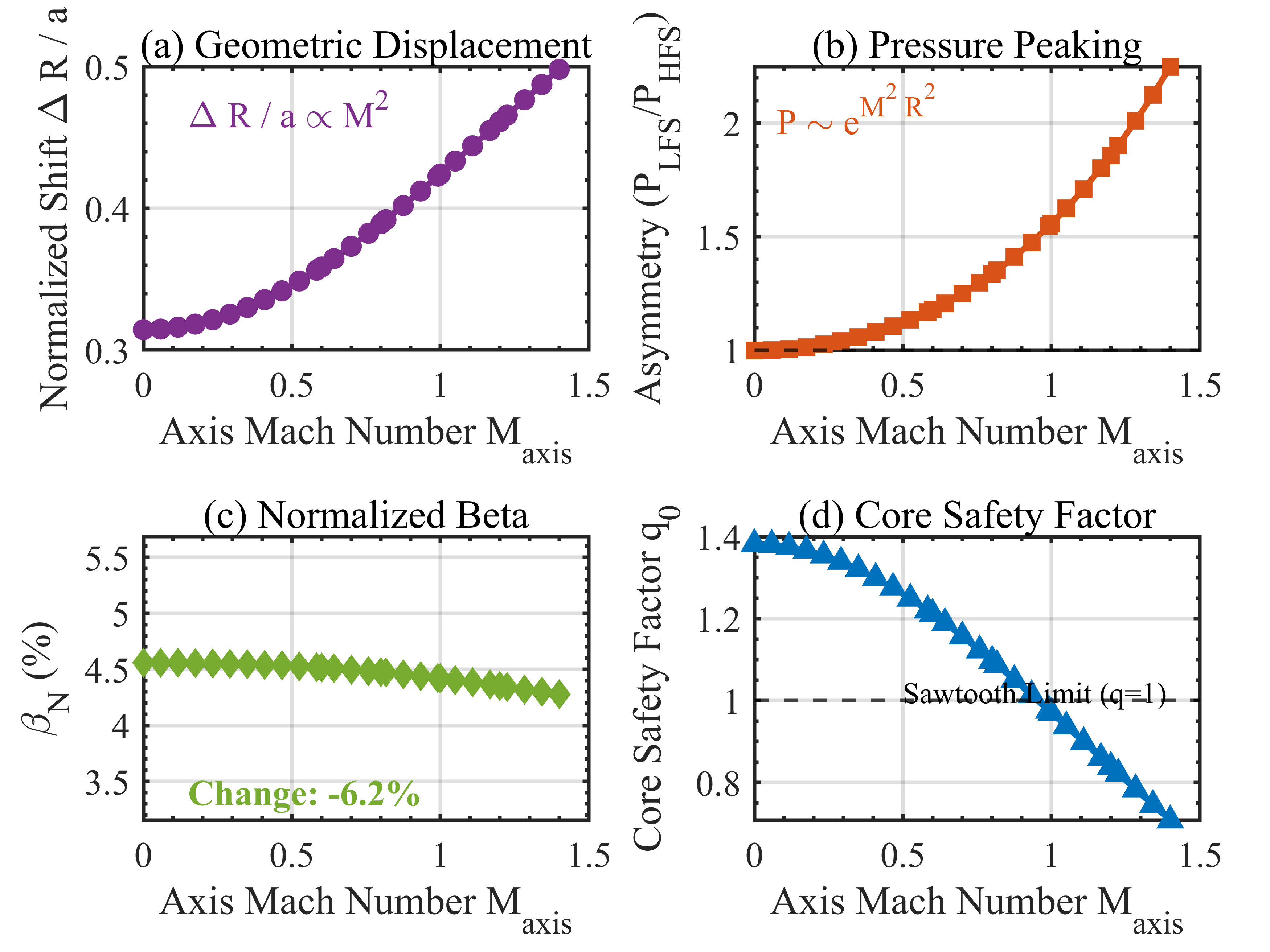}
    \caption{Impact of toroidal rotation on macroscopic equilibrium properties. (a) Normalized Shafranov shift $\Delta R/a$, exhibiting supra-linear growth with Mach number; (b) Thermodynamic asymmetry factor ($P_{LFS}/P_{HFS}$), indicating severe pressure pile-up; (c) Global normalized beta $\beta_N$, showing a slight reduction due to geometric reconfiguration; (d) Core safety factor $q_0$, dropping monotonically towards unity, highlighting the increased risk of sawtooth instabilities in sonic regimes.}
\end{figure}

\subsection{Mechanism of Topological Decoupling and Profile Modification}
To elucidate the microscopic origins of these macroscopic trends, Fig. 8 presents the radial profiles of key physical quantities across the midplane for rotation speeds up to $M=1.4$. The pressure profiles in Fig. 8a unveil a defining feature of strong rotation physics: the spatial decoupling between the magnetic topology and the thermodynamic mass center. In the static limit, the magnetic axis ($R_{mag}$, marked by hollow circles) and the pressure peak ($R_{Pmax}$, marked by solid diamonds) coincide. However, as the Mach number increases, the pressure peak migrates outward significantly faster than the magnetic axis. In the supersonic case ($M=1.4$), the pressure maximum is effectively ``detached'' from the magnetic axis, driven by the centrifugal potential $\sim \exp(M^2R^2)$ which overcomes the magnetic tension. As discussed by Waelbroeck\cite{Waelbroeck1996}, this decoupling creates an extremely steep pressure gradient at the LFS edge, which, while potentially conducive to transport barrier formation, poses a risk for ballooning mode instabilities.

This redistribution of mass necessitates a corresponding reconfiguration of the current system. The toroidal current density $J_\phi$ (Fig. 8b) does not increase uniformly; instead, it exhibits a pronounced asymmetric pile-up on the low-field side, forming a characteristic ``shoulder'' structure. This redistribution is a direct response to the flux compression discussed earlier. Simultaneously, the poloidal current function $F(\psi)$ (Fig. 8c) undergoes adaptive adjustment to maintain radial force balance against the expansive centrifugal stress, reflecting a dynamic shift in the plasma's diamagnetic/paramagnetic response. The cumulative effect of these changes is imprinted on the safety factor profile $q$ (Fig. 8d), which shifts downward globally. Crucially, the magnetic shear in the core region weakens and flattens as rotation increases. The combination of a low $q_0$ approaching unity and a weak-shear configuration constitutes a classic trigger for sawtooth crashes, as reviewed by Chapman\cite{Chapman2011} and modeled by Porcelli et al.\cite{Porcelli1996}, implying that supersonic rotation regimes may require auxiliary current drive to actively sustain the core safety factor. The ability of the VEQ-R solver to robustly capture these extreme profile distortions at $M=1.4$ without numerical instability demonstrates its applicability to the rigorous analysis of future high-performance burning plasmas.

\begin{figure}[H]
    \centering
    \includegraphics[width=0.85\linewidth, height=0.3\textheight, keepaspectratio]{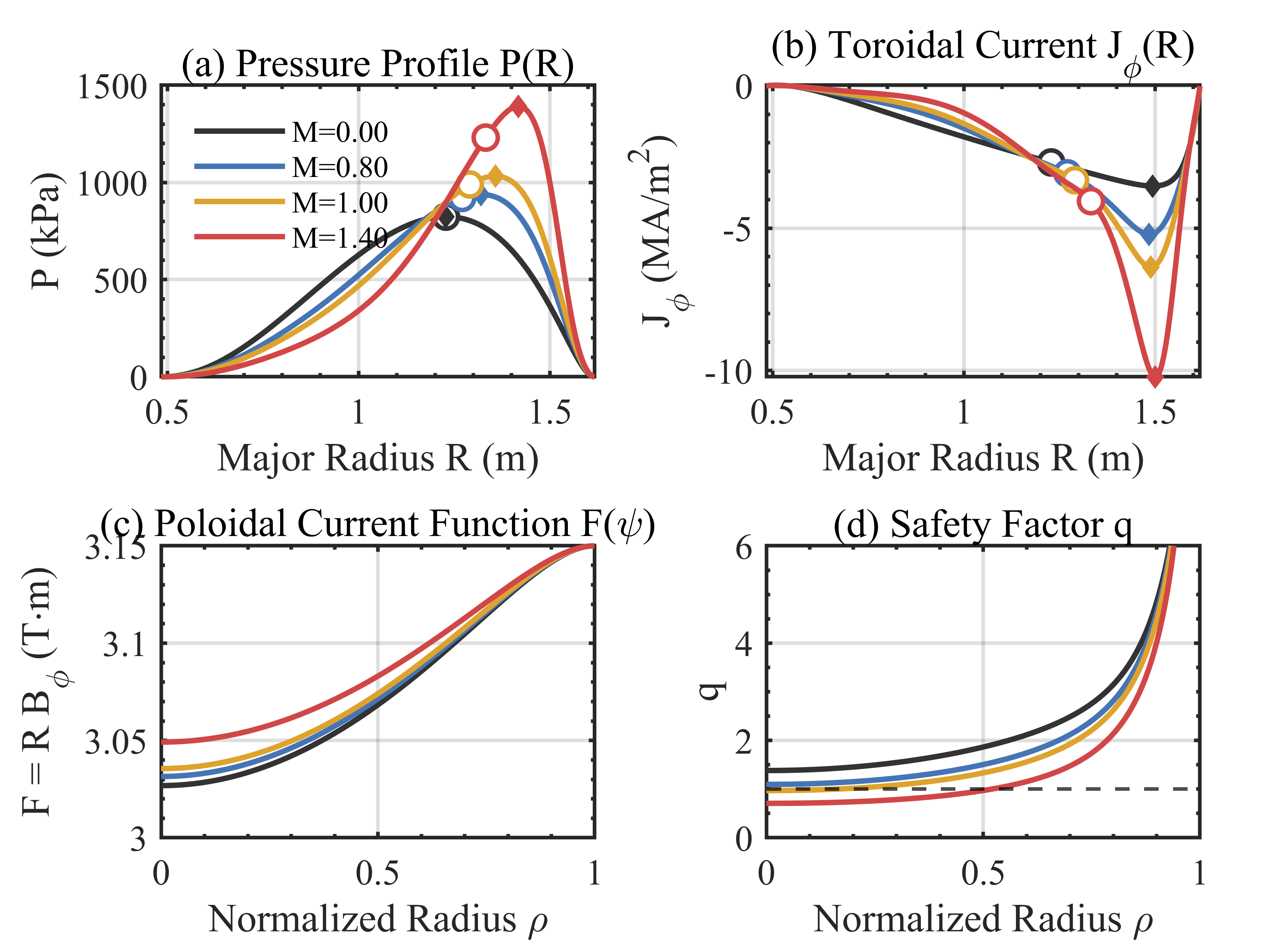}
    \caption{Modification of microscopic profiles by strong rotation (scanning from $M=0$ to $M=1.4$). (a) Pressure profiles $P(R)$ showing the spatial decoupling of the pressure peak (solid diamonds) from the magnetic axis (hollow circles); (b) Toroidal current density $J_\phi(R)$ exhibiting a pronounced ``shoulder'' on the low-field side; (c) Adaptive response of the poloidal current function $F(\rho)$; (d) Safety factor profiles $q(\rho)$, revealing the weakening of magnetic shear in the core region as rotation increases.}
\end{figure}

To visualize the structural origin of the pressure peak displacement shown in Fig. 8, we present a 2D contour comparison of the thermodynamic variables in Figure 9. This visualization clearly disentangles the thermal and inertial contributions to the equilibrium state. As observed in Fig. 9(b), the temperature contours (red solid lines) retain the characteristic `D-shape' of the magnetic flux surfaces. Although the center of the temperature profile shifts outward along with the magnetic axis (marked by the red cross), the isotherms remain congruent with the magnetic topology. This verifies that the numerical solver correctly preserves the imposed model assumption $T=T(\psi)$, ensuring thermodynamic consistency throughout the iteration.

In stark contrast, the density distribution in Fig. 9(c) exhibits a fundamental topological decoupling. The density contours are not only shifted but strongly compressed against the low-field side boundary, diverging significantly from the shape of the magnetic flux surfaces. This distortion is the direct manifestation of the centrifugal potential factor $\exp(M^2R^2)$, which acts to stratify the plasma mass toward larger major radii independent of the magnetic shear. Consequently, the macroscopic pressure shift observed in Fig. 9(a) is primarily driven by this inertial density pile-up rather than thermal gradients. This result highlights a critical feature of rapidly rotating STs: the isobaric surfaces (constant $P$) no longer coincide with the isothermal surfaces (constant $T$), creating a complex non-aligned gradient structure that may influence particle transport and stability boundaries.

\begin{figure}[H]
    \centering
    \includegraphics[width=0.85\linewidth, height=0.3\textheight, keepaspectratio]{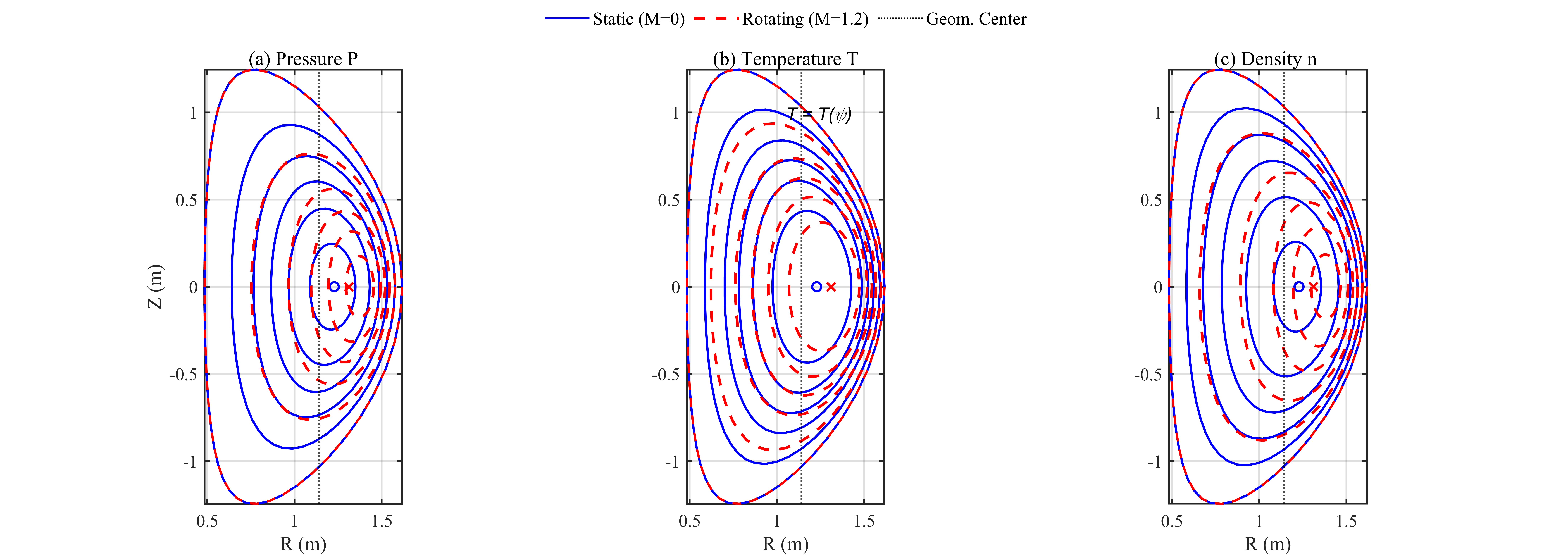}
    \caption{Comparison of 2D thermodynamic isocontours between the static ($M=0$, blue dashed lines) and transonic ($M=1.2$, red solid lines) equilibria. (a) Total Pressure $P$, exhibiting a significant outward shift; (b) Ion Temperature $T$, which remains a flux function $T(\psi)$ following the magnetic geometry; (c) Density $n = P/T$, showing severe centrifugal stratification towards the low-field side. The markers indicate the magnetic axis for the static (blue circle) and rotating (red cross) cases. Note that the density contours in (c) deviate significantly from the magnetic flux surfaces compared to the temperature contours in (b).}
\end{figure}

\subsection{Dependence on Aspect Ratio and Beta: Implications for Spherical Tori}
The distinct features of the VEQ-R solver—rapid convergence and geometric flexibility—make it uniquely capable of conducting extensive multidimensional parameter scans. While the previous sections analyzed a representative low-aspect-ratio configuration ($A=1.85$), we now extend the analysis to investigate the specific susceptibility of Spherical Torus (ST) plasmas to rotation compared to conventional tokamaks. Figure 10 summarizes these dependencies through a 2x2 scan of macroscopic equilibrium parameters.

First, we examine the impact of geometric compactness (Left Column: Figs. 10a \& 10c). Fixing the rotation at the sonic level ($M_{axis}=1.0$), we scanned the aspect ratio from the deep ST regime (1.3) up to the conventional range (3.0). The results in Fig. 10a uncover a strong non-linear dependency: low-aspect-ratio plasmas are significantly more susceptible to rotation-induced deformation. As $A$ decreases towards the ST limit of 1.3, the normalized axis shift $\Delta R/a$ nearly doubles compared to the conventional case. Concurrently, Fig. 10c reveals that this geometric compression exacts a toll on stability: the core safety factor $q_0$ drops significantly in the ST regime. This confirms that in compact devices, the steeper gradient of the centrifugal potential $R\Omega^2$ drives a stronger reconfiguration of the $q$-profile, potentially lowering the threshold for internal kink modes.

Second, we explore the interplay with plasma pressure (Right Column: Figs. 10b \& 10d). Fixing the geometry at our baseline ($A=1.85$), we scanned the normalized beta $\beta_N$. Fig. 10b demonstrates that the rotation-induced centrifugal shift adds constructively to the pressure-driven Shafranov shift, leading to a monotonic increase in total displacement. However, the most profound physical insight emerges from the response of the core safety factor $q_0$ (Fig. 10d). In static or slowly rotating plasmas, increasing $\beta_N$ typically drives a strong paramagnetic effect, forcing the toroidal current to concentrate near the magnetic axis and causing $q_0$ to drop rapidly—a linear degradation often leading to instability. Yet, under sonic rotation ($M=1.0$), VEQ-R reveals a distinct non-linear saturation mechanism. As observed in Fig. 10d, the decline of $q_0$ significantly flattens at high $\beta_N$ (change in slope). This stabilization occurs because the strong centrifugal potential $R^2\Omega^2$ effectively ``anchors'' the mass and pressure distribution towards the low-field side, establishing a dominant rotational equilibrium state. This centrifugal redistribution acts as a stiff geometric constraint that resists the further acute peaking of current density usually driven by increasing pressure. In essence, the rotation-induced ``centrifugal hold'' partially counteracts the pressure-driven ``paramagnetic squeeze,'' preventing $q_0$ from collapsing linearly and thereby preserving a higher stability margin than static theory would predict.

\begin{figure}[H]
    \centering
    \includegraphics[width=0.85\linewidth, height=0.3\textheight, keepaspectratio]{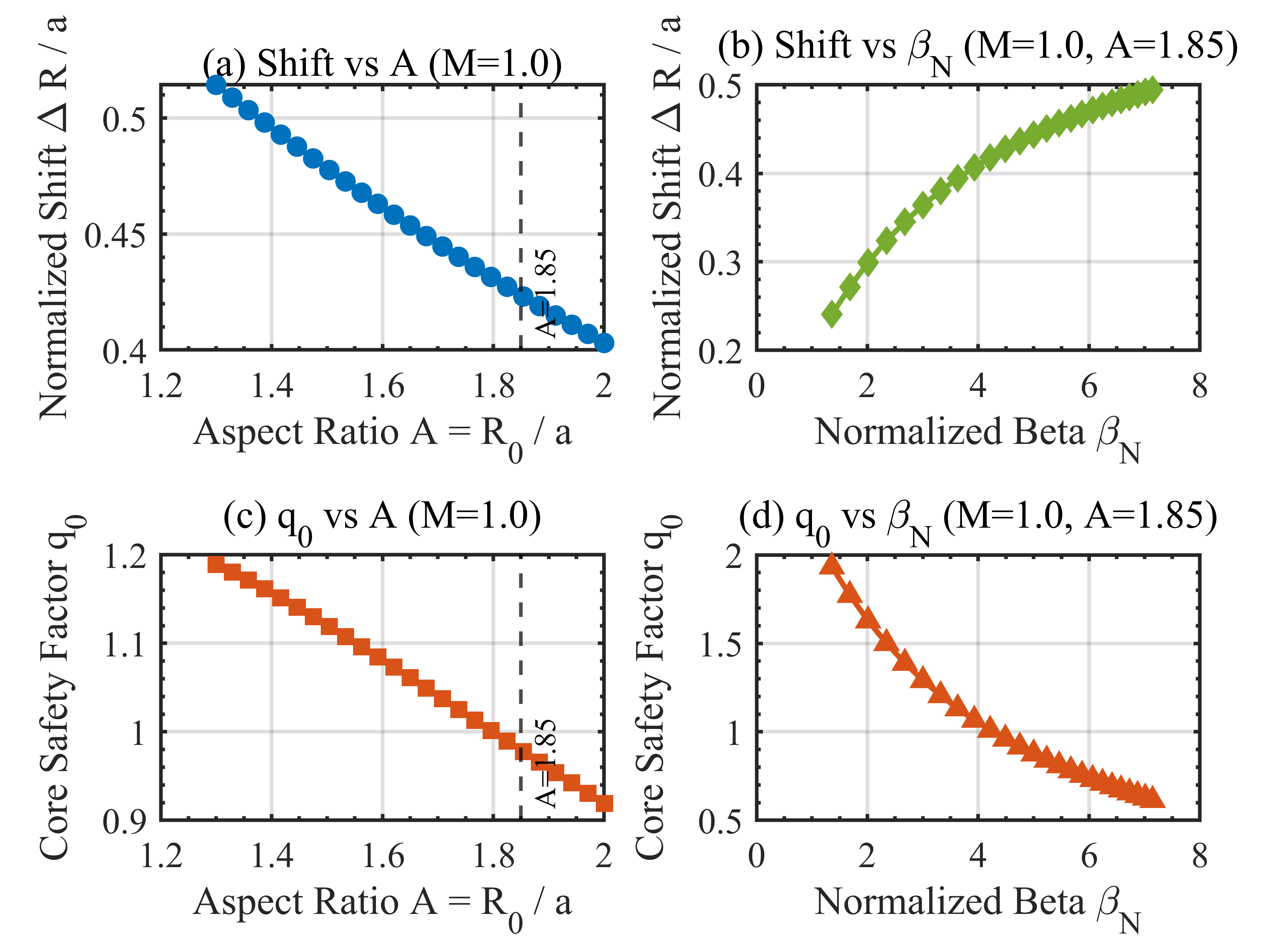}
    \caption{Multi-dimensional parameter scans revealing the enhanced sensitivity of Spherical Torus (ST) equilibria to rotation ($M_{axis}=1.0$) over an extended parameter space. Left Column: Dependence on Aspect Ratio $A$. (a) Normalized Shafranov shift $\Delta R/a$, showing drastic amplification as $A$ decreases into the ST regime; (c) Core safety factor $q_0$, illustrating the degradation of stability margins in compact geometries. The shaded region highlights the ST regime ($A < 2.0$), and the vertical dashed line marks the baseline $A=1.85$. Right Column: Dependence on normalized beta $\beta_N$ (at fixed ST-relevant geometry $A=1.85$). (b) Normalized Shafranov shift $\Delta R/a$, showing the constructive superposition of centrifugal and pressure-driven displacements; (d) Core safety factor $q_0$, exhibiting a non-linear saturation response where rotational flux compression competes with paramagnetic effects at high pressure.}
\end{figure}

\section{Conclusion}

In this work, we have addressed the persistent conflict between real-time computational capability and physical fidelity in the equilibrium reconstruction of strongly rotating plasmas. By introducing a 12-parameter Variational Moments Method (VEQ-R) powered by a novel Matrix-Kernel acceleration scheme, we have successfully bridged the gap that has long separated rapid control-oriented models from high-precision physics codes.

\subsection{Summary of Contributions and Physics Findings}
The cornerstone of this research is the development of VEQ-R, a solver that strikes an optimal balance between computational efficiency and geometric flexibility. By transforming the computationally expensive variational integration into pre-calculated algebraic matrix operations via the ``Matrix-Kernel'' technique, we achieved a convergence time of approximately 5 ms on a standard processor. This represents a speedup of three orders of magnitude compared to high-resolution Finite Difference Method (FDM) benchmarks.

Crucially, this speed does not compromise physical fidelity. Benchmarking confirms that VEQ-R reproduces magnetic flux surface geometries ($R, Z$) that are visually indistinguishable from high-resolution finite difference results. Quantitatively, the solver maintains relative errors for critical physical profiles—including pressure $P(R,Z)$, poloidal current function $F(\psi)$, and toroidal current density $J_\phi(R,Z)$—consistently within 1\%, while the safety factor $q(\psi)$ is captured with a relative error within 5\%.

Unlike traditional reduced models that often fail under extreme conditions, VEQ-R demonstrates exceptional robustness in the sonic rotation regime ($M \sim 1.0$). It accurately adapts its high-order spectral coefficients to capture ``non-rigid'' geometric distortions, successfully reproducing extreme features such as a 0.51 Normalized Shafranov shift $\Delta R/a$ and a thermodynamic asymmetry factor of 1.56. Furthermore, the solver distinguishes itself through its user-centric design. By allowing for the direct input of experimental physical profiles (e.g., Temperature and Rotation) rather than abstract flux functions, and by offering an inherently scalable spectral architecture, VEQ-R serves as a practical, high-throughput tool suitable for real-time control systems and massive integrated modeling workflows.

Beyond numerical acceleration, our parameter scans have uncovered subtle but critical stability implications of strong rotation. We identified a mechanism where centrifugal flux compression forces a concentration of the core current density, causing the safety factor $q_0$ to drop monotonically toward unity. This finding serves as a significant warning that strong rotation, despite its benefits for turbulence suppression, may inadvertently lower the threshold for $m/n=1/1$ internal kink modes and sawtooth crashes. Furthermore, we observed a complex performance trade-off: while centrifugal forces significantly enhance local pressure peaking, the necessary geometric reconfiguration results in a slight net reduction ($\sim 6.2\%$) of the global normalized beta. This suggests that rotation-induced confinement improvement is not a monotonic gain but requires a careful balance between local transport barriers and global magnetohydrodynamic stability limits.

\subsection{Future Outlook}
Building on this foundation, a strategic avenue for future research is the extension of this spectral framework into a multi-fluid MHD equilibrium solver\cite{XieMultiFluid}. This development is particularly critical for Spherical Tokamaks (STs) operating with advanced fuel cycles, such as Hydrogen-Boron ($p$-$^{11}$B) fusion\cite{Liu2024}. In such scenarios, the significant mass disparity between protons and boron ions, combined with the rapid rotation inherent to STs, induces profound centrifugal separation effects. Unlike single-fluid models, a multi-fluid description allows each species to develop a distinct density profile and, crucially, a distinct toroidal flow velocity field. Our high-order spectral method is uniquely suited to resolve the resulting poloidal asymmetries, where heavy impurities pile up on the low-field side while lighter fuel ions remain more centrally confined. By solving for the differential velocities of each fluid species, the extended solver will be capable of self-consistently reconstructing the complex multi-species current profiles arising from these separated flows. This capability will be essential for exploring the stability boundaries and current drive mechanisms in future aneutronic fusion reactors.

\section*{Acknowledgments}
The authors would like to thank Ruohan Zhang, Caixue Chen, Huibin Zhou, and Xinyu Liu for helpful discussions and kind assistance during this work. This research is supported by the National Natural Science Foundation of China (Grant Nos. 12435014 and 12475214), the National MCF Energy R\&D Program of China (Grant No. 2022YFE03090000), the Fundamental Research Funds for the Central Universities of Ministry of Education of China (Grant No. DUT25Z2536), and the Natural Science Foundation of Liaoning (Grant No. 2025-MSLH-157).



\begin{thebibliography}{99}

\bibitem{Rice2016}
Rice J.E. Experimental observations of driven and intrinsic rotation in tokamak plasmas. Plasma Phys. Control. Fusion 58, 083001 (2016).

\bibitem{Bondeson1994}
Bondeson A. and Ward D.J. Stabilization of external kink modes by resistive walls and plasma rotation. Phys. Rev. Lett. 72, 2709 (1994).

\bibitem{Hender2004}
Hender T.C., Buttery R.J. et al. Stability of neoclassical tearing modes in high performance JET plasmas. Phys. Plasmas 11, 4038 (2004).

\bibitem{Burrell1997}
Burrell K.H. Effects of $E \times B$ velocity shear and magnetic shear on turbulence and transport in magnetic confinement devices. Phys. Plasmas 4, 1499 (1997).

\bibitem{Maschke1980}
Maschke E.K. and Perrin H. Exact analytical solutions for the toroidal MHD equilibrium equation with mass flow. Plasma Phys. 22, 579 (1980).

\bibitem{Hameiri1983}
Hameiri E. Equilibrium and stability of rotating plasmas. Phys. Fluids 26, 230 (1983).

\bibitem{Lao1985}
Lao L.L., Stambaugh R.D. et al. Reconstruction of current profile parameters and plasma shapes in tokamaks. Nucl. Fusion 25, 1611 (1985).

\bibitem{Feng2024}
Feng X., Wu Z. et al. A new flux coordinates-based solver for fixed-boundary tokamak equilibrium with toroidal flow. Phys. Plasmas 31, 012505 (2024).

\bibitem{Chen2022}
Chen W., Ma Z., Zhang H., Zhang W. and Yan L. Free-boundary plasma equilibria with toroidal plasma flows. Plasma Sci. Technol. 24, 035101 (2022).

\bibitem{Chen2026}
C. X. Chen, H. S. Xie, and W. Kang, Comparison of Tokamak System Code Results for ENN's Spherical Torus with and without an Equilibrium Solver, submitted (2026).

\bibitem{Lao1981}
Lao L.L., Hirshman S.P. and Wieland R.M. Variational moment solutions to the Grad-Shafranov equation. Phys. Fluids 24, 1431 (1981).

\bibitem{Haney1989}
Haney S.W. and Freidberg J.P. Variational methods for the Grad-Shafranov equation. Phys. Fluids B 1, 1637 (1989).

\bibitem{Haney1995}
Haney S.W., Freidberg J.P. and Solomon C.J. A fast, user-friendly code for calculating magnetohydrodynamic equilibria. Comput. Phys. 9, 216 (1995).

\bibitem{Varadarajan1991}
Varadarajan V. and Miley G.H. Variational methods for studying tokamak plasma equilibria with arbitrary flow. Phys. Fluids B 3, 736 (1991).

\bibitem{Xie2026}
Xie H.S. and Li Y.Y. What Is the Minimum Number of Parameters Required to Represent Solutions of the Grad-Shafranov Equation? arXiv preprint arXiv:2601.02942 (2026).

\bibitem{Jardin2010}
S. Jardin, Computational Methods in Plasma Physics, CRC Press, 2010.

\bibitem{Hinton1985}
Hinton F.L. and Wong S.K. Neoclassical ion transport in rotating axisymmetric plasmas. Phys. Fluids 28, 3082 (1985).

\bibitem{Waelbroeck1996}
Waelbroeck F.L. Sonic flow and the distinct features of the equilibrium of rotating plasmas. Phys. Plasmas 3, 1047 (1996).

\bibitem{Boyd2001}
Boyd J.P. Chebyshev and Fourier Spectral Methods. 2nd ed. (Dover Publications, New York, 2001).

\bibitem{Canuto2006}
Canuto C., Hussaini M.Y., Quarteroni A. and Zang T.A. Spectral Methods: Fundamentals in Single Domains. (Springer, Berlin, 2006).

\bibitem{Broyden1965}
Broyden C.G. A class of methods for solving nonlinear simultaneous equations. Math. Comput. 19, 577 (1965).

\bibitem{Kelley1995}
Kelley C.T. Iterative Methods for Linear and Nonlinear Equations. (SIAM, Philadelphia, 1995).

\bibitem{Hansen1998}
Hansen P.C. Rank-Deficient and Discrete Ill-Posed Problems. (SIAM, Philadelphia, 1998).

\bibitem{Davis2006}
Davis T.A. Direct Methods for Sparse Linear Systems. (SIAM, Philadelphia, 2006).

\bibitem{Guazzotto2004}
Guazzotto L. et al. Flow and resistive wall mode stability in tokamaks. Phys. Plasmas 11, 604 (2004).

\bibitem{Belien2002}
Belien A.J.C. et al. FINESSE: Axisymmetric MHD equilibria with flow. J. Comput. Phys. 182, 91 (2002).

\bibitem{Chapman2011}
Chapman I.T. Controlling sawtooth oscillations in tokamak plasmas. Plasma Phys. Control. Fusion 53, 013001 (2011).

\bibitem{Porcelli1996}
Porcelli F., Boucher D. and Rosenbluth M.N. Model for the sawtooth period and crash. Plasma Phys. Control. Fusion 38, 2163 (1996).

\bibitem{XieMultiFluid}
Xie H.S., Li X.Y. et al. Development of a Reduced Multi-Fluid Equilibrium Model and Its Application to Proton-Boron Spherical Tokamaks. arXiv preprint arXiv:2602.09205 (2026).

\bibitem{Liu2024}
Liu M. et al. ENN's roadmap for proton-boron fusion based on spherical torus. Phys. Plasmas 31, 062507 (2024).


\end{thebibliography}
\end{document}